\documentclass[pdflatex,sn-mathphys-num]{sn-jnl}

\usepackage[table]{xcolor}
\usepackage{graphicx}%
\usepackage{multirow}%
\usepackage{amsmath,amssymb,amsfonts}%
\usepackage{amsthm}%
\usepackage{mathrsfs}%
\usepackage[title]{appendix}%
\usepackage{xcolor}%
\usepackage{textcomp}%
\usepackage{manyfoot}%
\usepackage{booktabs}%
\usepackage[ruled,vlined]{algorithm2e}
\usepackage{listings}%
\usepackage{anyfontsize}

\SetCommentSty{mycommfont}

\begin{document}

\title[Enhancing Collaborative Filtering-Based Course Recommendations by
Exploiting Time-to-Event Information with Survival Analysis]{Enhancing Collaborative Filtering-Based Course Recommendations by
Exploiting Time-to-Event Information with Survival Analysis}


\author*[1,2]{Alireza Gharahighehi}\email{alireza.gharahighehi@kuleuven.be}
\equalcont{These authors contributed equally to this work.}
\author*[1,2]{Achilleas Ghinis}\email{achilleas.ghinis@.kuleuven.be}
\equalcont{These authors contributed equally to this work.}

\author[1,2]{Michela Venturini}\email{michela.venturini@kuleuven.be}

\author[2,3]{Frederik Cornillie}\email{frederik.cornillie@kuleuven.be}

\author[1,2]{Celine Vens}\email{celine.vens@kuleuven.be}

\affil*[1]{\orgdiv{Department of Public Health and Primary Care}, \orgname{KU Leuven, Campus Kulak}, \orgaddress{\street{Etienne Sabbelaan 53}, \city{Kortrijk}, \postcode{8500}, \country{Belgium}}}

\affil[2]{\orgdiv{Itec, imec research group at KU Leuven}, \orgaddress{\street{Etienne Sabbelaan 51}, \city{Kortrijk}, \postcode{8500}, \country{Belgium}}}

\affil[3]{\orgdiv{Department of Linguistics}, \orgname{KU Leuven}, \orgaddress{\street{Etienne Sabbelaan, 53}, \city{Kortrijk}, \postcode{8500}, \country{Belgium}}}


\abstract{Massive Open Online Courses (MOOCs) are emerging as a popular alternative to traditional education, offering learners the flexibility to access a wide range of courses from various disciplines, anytime and anywhere. Despite this accessibility, a significant number of enrollments in MOOCs result in dropouts. To enhance learner engagement, it is crucial to recommend courses that align with their preferences and needs. Course Recommender Systems (RSs) can play an important role in this by modeling learners' preferences based on their previous interactions within the MOOC platform. Time-to-dropout and time-to-completion in MOOCs, like other time-to-event prediction tasks, can be effectively modeled using survival analysis (SA) methods. In this study, we apply SA methods to improve collaborative filtering recommendation performance by considering time-to-event in the context of MOOCs. Our proposed approach demonstrates superior performance compared to collaborative filtering methods trained based on learners' interactions with MOOCs, as evidenced by two performance measures on three publicly available datasets. The findings underscore the potential of integrating SA methods with RSs to enhance personalization in MOOCs.}

\keywords{recommendation systems, survival analysis, massive open online course, personalized learning, dropout}



\maketitle

\section{Introduction}\label{sec1}

Massive Open Online Courses (MOOCs) platforms offer a diverse range of online courses to learners around the globe, promoting equitable education by breaking down barriers related to geography and time. However, despite their significant advantages, many MOOC enrollments end in dropouts. Reports indicate that dropout rates for courses from renowned institutions such as MIT and Harvard can reach up to 90\%~\cite{chen2022systematic}. Dropouts may occur for various reasons, including accessing only the free parts of the courses, perceiving the course or topic as irrelevant, or lacking necessary competencies. This dropout information is crucial for modeling users' preferences and needs on MOOC platforms. Recommender Systems (RSs) are machine learning models that leverage users' past interactions to suggest the most suitable items to be recommended to the target user. Typically, RSs are divided into two main categories: Content-based filtering (CB) and collaborative filtering (CF). CB filtering RSs suggest items with features similar to those that the user has previously expressed interest in, while CF RSs predict users' preferences based on the similarities between users' and items' past interactions.

In a MOOC platform, a CF-based RS can be used to recommend courses based on users' previous enrollments. Although prior enrollments provide valuable data for modeling user preferences, they do not incorporate time-to-event information such as time-to-dropout or time-to-completion. Given the high dropout rates in MOOCs, incorporating time-to-event information can enhance the understanding of users' needs and preferences regarding courses. 

Survival analysis (SA) is a branch of statistics concerned with modeling the time until a particular event, such as death or machinery failure, occurs~\cite{clark2003survival}. A key aspect of survival data is that some events remain unobserved, known as censored data. Right-censoring, the most frequent type of censoring in SA, occurs when the target event is not witnessed during the follow-up period or if the instance is lost before the follow-up ends. The primary advantage of SA lies in its ability to use such partial data during the learning process by including instances with censored events which are often disregarded in classification and regression tasks. Our hypothesis is that incorporating the time-to-event (dropout or completion) is highly informative for modeling user preferences in MOOC recommendations, as it offers critical insights into students' engagement in MOOCs~\cite{room2021dropout}.

\begin{figure}
    \centering
    \includegraphics[width=0.9\linewidth]{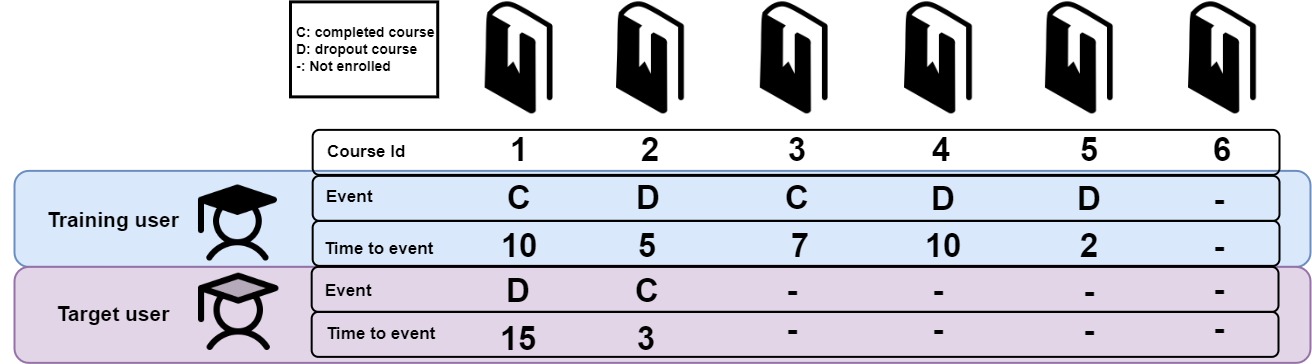}
    \caption{Illustration of time-to-event data in the context of MOOCs}
    \label{fig:mexample}
\end{figure}

In this paper\footnote{The source code is available at \url{https://anonymous.4open.science/r/mocc_cf_sa-85C9}}, we introduce a post-processing strategy utilizing SA to improve the effectiveness of CF techniques in the context of MOOCs. Our goal is to recommend courses that users are likely to enroll in with a high probability, and either complete swiftly or have a long time before dropout. This concept is demonstrated in Figure~\ref{fig:mexample}. Suppose a user, as shown in the figure, has enrolled in courses 1 to 5, completed courses 1 and 3, and dropped out of courses 2, 4, and 5. The target user, for whom we want to provide recommendations, has also enrolled in courses 1 and 2. Given the similar enrollments between the training and target users, standard CF methods would likely recommend courses 3, 4, and 5, suggesting they have higher predicted enrollment probabilities compared to course 6. However, these methods cannot effectively rank these courses among themselves. Ideally, courses expected to be completed fast or those with longer predicted dropout time should be ranked higher in the recommendation list. An SA model, trained on time-to-event information, will better quantify the ordering between the courses most likely to be enrolled in by the target user. The approach presented in this paper exploits time-to-event information (time-to-dropout or time-to-completion) to train an SA method and re-rank highly probable courses for enrollment based on their predicted time to dropout or completion.

The paper is organized as follows: relevant studies about MOOC recommendations and dropout prediction are described in Section~\ref{sec:LR}. Next, in Section~\ref{sec:method}, we illustrate our proposed approach, explaining the collaborative filtering task, the time-to-event prediction task and the final post-processing step to generate the final recommendation lists. In Section~\ref{sec:ES}, we describe three publicly available datasets and the experimental setup in designing and testing the proposed approach. Next, in Section~\ref{sec:result} we present and discuss the results of comparing our proposed approach against some CF methods on these three datasets. Finally, we conclude and outline some directions for future research in Section~\ref{sec:con}.

\section{Related work}
\label{sec:LR}
\subsection{MOOC recommendation}

Various types of RSs have been utilized in the context of MOOCs, including collaborative, knowledge-based, and content-based filtering. Among these, CF RSs have been extensively applied, either individually or in combination with other types, since they do not require item or user metadata to generate recommendations and can rely solely on learners' logs~\cite{uddin2021systematic}. Numerous studies have applied CF for MOOC recommendations, with nearest neighbors~\cite{fu2015undergraduate,he2017design,lu2019research,yang2014peer,song2016research,pang2018adaptive,yin2020mooc} and matrix factorization~\cite{wu2021collaborative,gharahighehi2023extending,chao2019collaborative} approaches being the most popular. Although time-related information provides relevant insights into learners' preferences and needs in MOOCs, few studies have incorporated this data into their MOOC recommendations. For instance, one study used learners' dwell time on the MOOC page in edX\footnote{\url{https://www.edx.org/}} to provide personalized recommendations~\cite{pardos2017enabling}. Another similar study~\cite{tang2017personalized} applied a time-augmented Recurrent Neural Network (RNN) to consider the amount of time learners spent on each course page for making personalized recommendations in edX. In our previous study~\cite{gharahighehi2023extending}, we demonstrated that SA can improve the performance of a specific RS, namely Bayesian Personalized Ranking (BPR), when the predictions of a SA method, trained based on time to dropouts, are embedded in the BPR algorithm. While SA based on time-to-dropout improved the quality of recommendations, it has only used in a specific algorithm, namely BPR.

While using time information has proven to have a positive effect on RS performance, to our knowledge, time-to-event data, such as time-to-completion and time-to-dropout, have not been utilized to provide more informed recommendations, specifically in CF RSs.

\subsection{Time-to-event prediction in MOOCs}
The task of dropout prediction in the context of MOOCs has been mainly modeled as a classification task~\cite{dalipi2018mooc, chen2022systematic}. While in these studies the task was predicting the event of dropout, the authors ignored the time information in their predictions. SA can be used to incorporate the time information in modeling dropout in MOOCs and there are some promising examples in the literature. The authors in~\cite{gitinabard2018your} used SA, specifically Cox proportional hazards method, to model dropout risk in the context of MOOCs and unveil social and behavioral features impacting the outcome. Xie~\cite{xie2019modelling} utilized survival analysis to examine the hazard function of dropout, employing the learner's course viewing duration on a course in MOOCs. Labrador et al.~\cite{labrador2019survival} specified the fundamental factors attached to learners' dropout in an online MOOC platform using Cox proportional hazard regression. Wintermute et al.~\cite{wintermute2021survival} applied a Weibull survival function to model the certificate rates of learners in a MOOCs platform, assuming that learners “survive” in a course for a particular time before stochastically dropping out. In~\cite{pan2022survival} a more sophisticated SA deep learning approach was proposed to tackle volatility and sparsity of the data, that moderately outperformed the Cox model. Masci et al.~\cite{masci2024modelling} applied shared frailty Cox models to model dropout of students who enrolled in engineering programs. 


Although SA has been applied to model dropout in MOOCs, to the best of our knowledge, it hasn't been used to model user preferences and needs in MOOC recommendations. The research gap that we aim to fill is to investigate the merits of SA to model time-to-events in the context of MOOCs, specifically time-to-dropout and time-to-completion, and use it to enhance the performance of typical CF RSs. 

\section{Methodology}
\label{sec:method}

\subsection{Problem formulation}

In recommendation tasks there are two main sets of entities, the users, who receive the recommendations, and the items, which can be recommended to the users. Let $U = \{u_{1}, u_{2}, ..., u_{m}\} $ and $I = \{i_{1}, i_{2}, ..., i_{n}\}$ be two finite sets, representing users and items, respectively. The already known interactions between such items and users are stored in an interaction matrix $\mathbf{M}$, which in the context of our study on MOOCS can contain tuples where the first element in the tuple contains the time to the event, and the second element of the tuple is the event between the user and course (``c" completed or ``d" dropout):

\begin{equation}
\label{eq:hyperdef}
 M_{ui} = \left\{ 
  \begin{array}{l l}
    (t_{ui},c), & \quad \mbox{if user $u$ completed course $i$}\\
    (t_{ui},d), & \quad \mbox{if user $u$ dropout from course $i$}\\
    0, & \quad \mbox{if user $u$ hasn't enrolled in course $i$}.
  \end{array} \right.
\end{equation}

\noindent where $t_{ui}$ is time-to-completion or time-to-dropout for user ``u" and item ``i". To represent learners enrolments in MOOCs, we consider enrolment matrix $E$ by binarizing the interaction matrix $M$. The task of MOOC recommendation is to provide a ranked $top@k$ recommendation list, i.e., the first $k$ items in the ordered list, to each user.

\subsection{Collaborative filtering}
\label{method:cf}
The task of a CF-based RS is to model user preferences over unseen items and generate ranked lists of recommendations using a sparse interaction matrix between users and items. CF RSs either form neighborhoods around users or items (UKNN or IKNN\footnote{User- or Item-based K Nearest Neighbors}) or learn latent features (e.g. SVD\footnote{Singular Value Decomposition} and NMF\footnote{Non-negative Matrix Factorization}) to infer preferences. In the context of MOOCs, the RS is trained on the enrollment matrix, which contains user enrollments. Once trained, the CF-based recommendation system can predict the missing values in the matrix, i.e., the courses that learners haven't yet enrolled in, thereby reconstructing the entire matrix. This allows the system to identify the MOOCs that learners are most likely to enroll in, and the courses will then be ranked based on these predictions with the $top@k$ courses recommended to the learner.

\subsubsection{Memory-based collaborative filtering}
User-based and item-based KNN (UKNN and IKNN) are memory-based CF methods that infer missing interactions between users and items by leveraging the data of neighboring users or items. UKNN and IKNN predict missing values in the interaction matrix by calculating a weighted average of the scores from similar users or items. The weights assigned to each neighbor represent the similarity between their interaction vector and that of the target user or item.

\subsubsection{Model-based collaborative filtering}
Model-based CF RSs learn latent features for items and users, and then use these features to construct the interaction matrix. For example, Pure Singular Value Decomposition (SVD)~\cite{cremonesi2010performance} and Non-negative Matrix Factorization (NMF)~\cite{cichocki2009fast} are CF RSs that decompose the interaction matrix into two low-rank matrices for users and items. In NMF, the user and item learned matrices contain only non-negative values. Given $\mathbf{P_{u}}$ and $\mathbf{Q_{i}}$ as the learned latent features of users and items respectively, the enrolment matrix can be reconstructed by multiplying $\mathbf{P_{u}}$ and $\mathbf{Q_{i}}$. 

\subsection{Survival Analysis}
\label{method:sa}
\subsubsection{Defining time-to-event and censoring}
\label{sec:survival_definitions}
In this context, we can define the time-to-event variable as the number of days elapsed between a users' first and last interactions with a given course. The definition of censoring and event times is dependent on whether the event of interest is course dropout or completion. For example, if the time-to-event variable of interest is course completion, then event times are defined as the total number days elapsed between a student's first and last interactions for a course they have successfully completed, while students who have not yet completed that course at their last interaction are considered to be censored.

Survival data contains two key components: a time $Y$ which denotes the time an individual was followed up for, and a binary event variable $\delta$ which denotes whether $Y$ corresponds to an event time when the event of interest occurred, or a censoring time where the individual was last observed without the event having occurred. 
Using the definitions from equation \ref{eq:hyperdef}, if the event of interest is defined as course completion, the tuple $(t_{ui},c)$ would correspond to a user who has experienced the event while the tuple $(t_{ui},d)$ corresponds to a user who is censored. If the event of interest is defined as course dropout, then the opposite is true where $(t_{ui},d)$ corresponds to a user who has experienced the event while $(t_{ui},c)$ corresponds to a censored user. Additionally, a set of covariates $X$ which could be predictive of a user's likelihood of successfully completing a course is often available on both the user and course levels. These covariates can be used as features in various SA models, such as like Regularized Cox Proportional Hazards Models (CoxNet), Gradient Boosted Ensembles (XGB), and Random Survival Forests (RSF), to predict the time to dropout or completion.

\subsubsection{Survival analysis definitions}
The Cox Proportional Hazards (CPH)~\cite{cox1972regression} is a semi-parametric method for estimating the hazard function $h(t,x)$ which measures the instaneous risk of experiencing the event at time $t$ given that the individual has not yet experienced it at time $t$. The hazard function can also be expressed as  $h(t)=\frac{d}{dt}H(t)$ where $H(t)$ is the Cumulative Hazard Function. While $H(t)$ does not have intuitive interpretation, the Survival Function $S(t)=P(T>t)=1-F(t)$ which denotes the probability that an individual does not experience the event before time $t$ can be expressed as $S(t)=\exp(-H(t))$. The CPH model is then defined as:

\begin{equation}
\label{eq:cph}
h(t,X_i) = h_0(t)\exp(X_i^T\beta)
\end{equation}

\noindent where $h_0(t)$ corresponds to a baseline hazard function which is common for all individuals and $\exp(X^T_i\beta)$ serves as a multiplicative factor affecting that baseline hazard based on an individuals covariates. In terms of estimation, the baseline hazard $h_0(t)$ is treated as a nuisance factor while the coefficients of $\beta$ are the main parameters of interest. If we define $T_1<\dots<T_J$ as the $J$ ordered distinct event times and assume there are no ties in event times, it can be shown~\cite{cox1975partial} that estimation for $\beta$ can be achieved by maximizing the log-partial likelihood:



\begin{equation}
\label{eq:cph_like}
LL(\beta) = \sum_{j=1}^N \delta_j \Big[X_j^T\beta - \log\Big(\sum_{i\in R_j}\exp(X_i^T\beta)\Big)\Big]
\end{equation}

\noindent where $X_j$ corresponds to the covariates of the individual who experienced the event at time $T_j$, while $R_j$ corresponds to the set of individuals still at risk of experiencing the event at time $T_j$. 

\subsubsection{Regularized Cox Proportional Hazards Model}

CoxNet~\cite{simon2011regularization} combines the well known $\ell_1$ lasso and $\ell_2$ ridge penalties on the coefficients of the CPH model in an elastic-net~\cite{zou2005regularization} fashion to introduce sparsity in high-dimensional problems and avoid overfitting. Given a set of covariates $p$, equation \ref{eq:cph_like} is modified to the corresponding objective function:

\begin{equation}
\arg \max_\beta \quad LL(\beta) - \alpha \Big( r\sum_{k=1}^p|\beta_k| + \frac{1-r}{2}\sum_{k=1}^p\beta_k^2\Big)
\end{equation}

\noindent where $r\in (0,1)$ controls the relative weight of the $\ell_1$ and $\ell_2$ penalties while $\alpha \in(0,1)$ controls the overall shrinkage.

\subsubsection{Gradient Boosted Ensembles}
Gradient Boosting is a common framework for predictive modeling which uses an ensemble of weak learners~\cite{Friedman2001}. In the context of SA, ~\cite{ridgeway_state_1999} proposed replacing the linear regression component of equation \ref{eq:cph_like} with a boosted ensemble of regression based estimators $f(x)$ to maximize the log-partial likelihood:
\begin{equation}
    LL(\beta) = \sum_{j=1}^N \delta_j \Big[f(x) - \log\Big(\sum_{i\in R_j}\exp(f(x)\Big)\Big]
\end{equation}

\noindent where a popular implementation of boosted Cox models uses regression trees for the weak-learners~\cite{polsterl2020scikit}.

\subsubsection{Random Survival Forests}
Random Survival Forests (RSF)~\cite{Ishwaran2008} are an extension of Random Forests~\cite{Breiman2001} to specifically model time-to-event outcomes with censored observations where individual trees within an RSF are grown to maximize the survival difference between nodes. The most common splitting criterion makes use of the log-rank test~\cite{Bland2004} between the resulting nodes. Once a tree is grown, the Cumulative Hazard Function within each terminal node $h$ is calculated using the non-parametric Nelson-Aalen estimator as:

\begin{equation}
H(t|x_i)=\hat{H}_h(t) = \sum_{t_{j,h}\le t} \frac{d_{j,h}}{R_{j,h}}
\end{equation}

\noindent where $d_{j,h}$ is the number of events at time $t_{j,h}$ and $R_{j,h}$ is the number of individuals at risk at time $t_{j,h}$. The ensemble estimator for the Cumulative Hazard Function is then obtained by averaging all the individual trees. 


\subsubsection{Model Fitting and Interpretation}

The survival models can be fit using the known interactions of users and courses in a MOOC dataset. Each training instance is defined on an observed user-course interaction and the covariates for that instance can consist of both user-level and course-level information for the given user-course pair (the applied features in the experiments are discussed at the end of Section~\ref{sec:ES_d}). The target event times and indicators can be constructed based on whether the event of interest is course dropout or course completion as described in section \ref{sec:survival_definitions}. To avoid biased predictions based on the overall duration of a course, the time-to-event variable can be normalized within each course such that the minimum and maximum days elapsed between the first and last interactions of users within that course are the same across all courses in the database while information on the duration can be included as a covariate in the model. A prediction set of instances consisting of unseen user-course interactions can be constructed in a similar fashion as in the training stage. Now, each row corresponds to user-course pairs which are unobserved with the same user-level and course-level covariates as in the training stage. The survival model can now be used to predict a risk score for each of these unobserved interactions. 

The key distinction between modeling time-to-dropout and time-to-completion lies in the interpretation of the risk predictions. When modeling time-to-completion, a higher risk score for a user between course $A$ and course $B$ indicates that the student is likely to complete course $A$ faster, relative to the average student, compared to course $B$. A recommender would prioritize courses where a user has a high risk score, which corresponds to courses that the user is more likely to complete quickly. Conversely, when modeling time-to-dropout, a higher risk score for a user between course $A$ and $B$ means that the student is likely to drop out of course $A$ faster, relative to the average student, compared to course $B$. Therefore, a recommender would prioritize courses where the user has a low risk score, which corresponds to courses that the user is less likely to drop out of quickly and more likely to engage with for a longer duration. 

The two models thus capture two distinct user behaviors: how quickly they will complete a course and how quickly they will drop out of a course.  Instead of choosing between the time-to-completion and time-to-dropout models, predictions from both models can be utilized to identify courses that a user has both a high probability of completing quickly and a low probability of dropping out quickly. This can be achieved by aggregating the ranks of courses based on their dropout risk scores from lowest to highest and their completion risk scores from highest to lowest, and then ordering the courses based on the average ranks from these two lists.

\subsection{Re-ranking}
The main idea of this paper is to enhance the performance of CF-based RSs using predictions from a SA model trained on time-to-event data in the context of MOOCs. As discussed in Section~\ref{method:cf}, CF-based RSs are designed to rank the courses that learners are most likely to enroll in next (Step~1 in Algorithm~\ref{alg:alg}), based on their previous enrollments. By incorporating predictions from SA methods, which are trained on time-to-dropout ($\hat Y_{d}$ in Algorithm~\ref{alg:alg}) or time-to-completion data ($\hat Y_{c}$ in Algorithm~\ref{alg:alg}), or their aggregated ranked list ($L_{SA_{CD}}$ in Algorithm~\ref{alg:alg}), the initial list generated by a CF-based RS can be re-ranked. This re-ranking prioritizes courses that, among the initially ranked ones with the CF recommender, have shorter predicted time-to-completion or longer predicted time-to-dropout. The entire concept is illustrated in Algorithm~\ref{alg:alg}.

\begin{algorithm}[tb]
\caption{Enhancing Course Recommendations Using Survival Analysis}
\label{alg:alg}

\KwIn{user-item interaction matrix $M$, initial list length $l$, final recommendation list length $k$}
\KwOut{top@k recommendation list for each user}

\BlankLine
\textbf{Step 1: Collaborative filtering} \\
$E \gets binarize(M)$ \\
$\hat E \gets CF.fit\_predict(E)$ \\
$ L_{CF} \gets rank(\hat E, l)$ \tcp*[h]{Rank courses based on CF predictions and keep the first $l$ for each user} 
\BlankLine
\textbf{Step 2: Survival analysis} \\
$X \gets get\_features(M)$ \\
$\hat Y_{c} \gets SA_{c}.fit\_predict(X, M)$ \tcp*[h]{SA model based on time-to-completion}
$\hat Y_{d} \gets SA_{d}.fit\_predict(X, M)$ \tcp*[h]{SA model based on time-to-dropout}
$ L_{SA_{C}} \gets rank(\hat Y_{c})$ \tcp*[h]{Rank courses based on time-to-completion SA predictions}
$ L_{SA_{D}} \gets rank(\hat Y_{d})$ \tcp*[h]{Rank courses based on time-to-dropout SA predictions}
$ L_{SA_{CD}} \gets aggregate\_rank(L_{SA_{c}}, L_{SA_{D}})$ \\
\BlankLine
\textbf{Step 3: Re-ranking} \\
$top@k = re\_rank(L_{CF}, L_{SA_{(C/D/CD)}})$ \\

\BlankLine
\textbf{Return} $top@k$

\end{algorithm}

\section{Experimental design}
\label{sec:ES}
\subsection{Datasets}\label{sec:ES_d}
Publicly available datasets generated from MOOCs are scarce and most of them are described by Lohse et al. in~\cite{lohse2019surveying}. To evaluate our approach, we used three widely recognized publicly available datasets: XuetangX~\cite{feng2019understanding}, KDDCUP~\cite{feng2019understanding}, and Canvas~\cite{canvas}. Both the KDDCUP and XuetangX datasets are anonymized and provided by the XuetangX platform\footnote{\url{https://www.xuetangx.com/}}. The Canvas dataset contains de-identified data from Canvas Network\footnote{\url{https://www.canvas.net/}} open courses from January 2014 to September 2015. Table~\ref{tab:data} describes the three preprocessed publicly available MOOC datasets that were used to evaluate the proposed approach.
The raw JSON files containing logs of all interactions an individual had with a course for the XuetangX and KDDCUP datasets were processed to extract the first and last interactions a user had with a given course. The time-to-event variable was defined as the difference between the dates of these actions and binary indicators denoting whether a user dropped out or completed the course were provided. The Canvas dataset was in tabular format and already contained that information. In all three datasets, the time-to-event variable was normalized within each course such that the minimum and maximum days elapsed between the first and last interactions of users within that course were the same for each course. 

Due to the lack of consistent high-quality metadata at both the student and course levels across the three datasets—such as age, education, or course descriptions—the covariates for the models were kept relatively simple. For each user, the number of courses taken, percentage of courses completed, and average completion and dropout times in the training set were included as features in the SA models. Additionally, for each student-course pair, the student level user-item interaction vector containing all enrollment and time-to-event data for that student in the training-set, as well as the course level item-user interaction vector containing all enrollments and time-to-event data students had with that course in the training set were included after performing dimensionality reduction using Principal Components Analysis and retaining the components containing 80\% of the total variance. 

\begin{table*}
\centering
  \caption{Datasets descriptions}
    \label{tab:data} 
   
  \begin{tabular}{lccc}
    \toprule
    & XuentangX&KDDCUP&Canvas\\
    \midrule
    Number of Users &2417&1944&959\\
   Number of Items &246&39&193\\
   Sparsity &$95.5\%$&$87.1\%$&$95.4\%$ \\
   Avg number of completed courses per user&4.6&4.8&4.3\\
   Avg number of dropout courses per user&5.9&4.4&4.5\\

  \bottomrule
\end{tabular}
\end{table*}

\subsection{Experimental setup}
\label{setup}
Before processing the datasets the cold-start users and courses that have less than 5 interactions where at least three of them should be course completion, were dropped. Then each dataset was split into three disjoint sets: training, validation and test sets. Test and validation sets contained three (at least one completed course) and one (either completed or dropout course) interaction per user respectively. The rest of interactions were used for training. The validation set was used to tune the hyperparameters (the details about hyperparameter tuning is reported in Appendix~\ref{hyper}). 


The five-fold cross-validated concordance index (C-Index) on the training set was used to tune and evaluate the performance of SA methods. The C-index can be seen as a generalization of the Area Under the Curve (AUC) in classification models, particularly when dealing with survival data that includes censored information. The metric essentially evaluates whether individuals with higher risk scores experience the event faster than those with lower risk scores~\cite{Longato2020}. Similar to AUC, the C-index ranges from 0 to 1. Additionally, two variants of the Normalized Discounted Cumulative Gain (NDCG) were considered to assess the final recommendations. NDCG is a rank-sensitive evaluation measure that penalizes recommendation scores if relevant items appear lower in the list. Apart from the regular NDCG, and in order to incorporate time-to-event information, we introduced a variant (NDCG-t), where relevance scores are linearly decayed based on either the maximum time-to-dropout for dropout courses or the minimum time-to-completion for completed courses. This approach prioritizes courses with longer dropout times or shorter completion times as more preferred.

\subsection{Competing approaches}
\label{baselines}
For each step mentioned in Algorithm~\ref{alg:alg}, different competing methods are applied. For the first step, we selected the CF baselines based on the results of the award winning paper~\cite{dacrema2019we}, which showed that simple CF RSs such as memory-based approaches (UKNN and IKNN), and SLIM outperform more recent complex deep neural network based approaches. Therefore, the following baselines are selected:

\begin{itemize}
    \item \textbf{UKNN} and \textbf{IKNN}: user- and item-based KNN~\cite{sarwar2001item,lops2011content} are memory-based CF methods that impute missing interactions between users and items based on the interactions of neighbor users/items. 
    \item \textbf{SVD}: Singular Value Decomposition (SVD)~\cite{cremonesi2010performance} can be applied to decompose the interaction matrix to two low-rank matrices for users and items. 
    \item \textbf{NMF}: Non-negative Matrix Factorization (NMF)~\cite{cichocki2009fast} is similar to SVD but the learned user and item matrices contain non-negative values.
    \item \textbf{WRMF}: weighted regularized matrix factorization (WRMF)~\cite{pan2008one} is a model-based CF method that utilizes the alternating-least-squares optimization algorithm to learn its parameters.
    \item \textbf{EASE}: Embarrassingly Shallow Autoencoders (EASE)~\cite{steck2019embarrassingly} is a linear collaborative filtering model based on shallow auto-encoders~\cite{cheng2016wide}.
    \item \textbf{SLIM}: Sparse LInear Method (SLIM)~\cite{ning2011slim} is a method that learns the sparse aggregation coefficient square matrix using the optimization problem regularized with L1 and L2 norms.
\end{itemize}

For the second step in Algorithm~\ref{alg:alg}, the following SA methods are considered:
\begin{itemize}
    \item \textbf{CoxNet} is a penalized variant of the Cox Proportional Hazards Model.
    \item \textbf{RSF} is random forest extension to survival or time-to-event outcomes.
    \item \textbf{XGB} is a boosting method using regression trees as base learners with a cox partial likelihood.
\end{itemize}

Finally, for the re-ranking step (the third step in Algorithm~\ref{alg:alg}), we followed three options to include SA predictions, re-ranking based on time-to-completion (re-ranking based on $ L_{SA_{C}}$), time-to-dropout (re-ranking based on $ L_{SA_{D}}$) and their combined ranks (re-ranking based on $ L_{SA_{CD}}$).

\section{Results and Discussion}
\label{sec:result}
Five-fold cross-validation on the training set was used to select the best parameters for each survival method to model time-to-completion or time-to-dropout. The cross-validated C-index  for each dataset and method is reported in Table~\ref{tab:SA} with the optimal parameters in Table~\ref{tab:hyper}. $XGB$ outperforms both $CoxNet$ and $RSF$ in terms of the C-index across all three datasets, for both time-to-completion and time-to-dropout prediction tasks. Except for the case where $XGB$ is applied to the Canvas dataset, SA methods trained on time-to-dropout generally perform better than those trained on time-to-completion. This suggests that time-to-dropout is more informative when modeling time-to-event in MOOCs. We chose $XGB$ to model time-to-event in our experiments due to its better C-index based on 5-fold cross-validation compared to $CoxNet$ and $RSF$. 

Table~\ref{tab:rerank_full} shows the results of applying several CF-based RSs and the proposed post-processing approaches on three MOOCs datasets, evaluated using two variants of the NDCG measure. In this table, the `Baseline' column includes CF RSs performance without post-processing with SA models predictions. `+D', `+C,' and `+DC' in the table stand for post-processing based on time-to-dropout, time-to-completion, and their combination, respectively. The best-performing approach in each dataset and for each measure is represented with the underlined numbers.

For the first step in Algorithm~\ref{alg:alg}, among the CF RSs, SLIM performs best for XuetangX and KDD, while UKNN is the top performer for Canvas according to both evaluation measures. As shown in the table, all three post-processing approaches perform better compared to the corresponding CF-based RS baseline. Post-processing based on both time-to-dropout and time-to-completion (`+DC') performs better in most cases compared to post-processing based on only one type of event, which implies that using SA predictions based on both events, i.e., dropout and completion, are more effective to model user preferences and needs.


\begin{table*}
\centering
  \caption{Survival analysis methods comparisons w.r.t. C-index}
    \label{tab:SA} 
  \begin{tabular}{@{}lcccccc@{}}
\toprule
       & \multicolumn{2}{c}{XuetangX}                                        & \multicolumn{2}{c}{Canvas}                                   & \multicolumn{2}{c}{KDD}                                      \\ \midrule
       & \multicolumn{1}{l}{Dropout} & \multicolumn{1}{l}{Completion} & \multicolumn{1}{l}{Dropout} & \multicolumn{1}{l}{Completion} & \multicolumn{1}{l}{Dropout} & \multicolumn{1}{l}{Completion} \\ \midrule
Coxnet & 0.7119                      & 0.7117                         & 0.7355                      & 0.7355                         & 0.7061                      & 0.6885                         \\
RSF    & 0.7223                      & 0.7064                         & 0.7827                      & 0.7722                         & 0.7475                      & 0.7148                         \\
XGB    & \textbf{0.7479}                      & \textbf{0.7269}                         & \textbf{0.7956}                      & \textbf{0.8079}                         & \textbf{0.8083}                      & \textbf{0.7309}                         \\ \bottomrule
\end{tabular}
\end{table*}
\newcommand{\cg}{\cellcolor{gray!20}}

\begin{table*}
\caption{Re-ranking with XGb for all baseline reccomenders}
\label{tab:rerank_full}
\centering
\scalebox{0.79}{
\begin{tabular}{cllcccc|cccc} 
\toprule
\multicolumn{1}{l}{}             &                             &                                    & \multicolumn{4}{c}{\textbf{ndcg}}         & \multicolumn{4}{c}{\textbf{ndcg-t}}                             \\ 
\midrule
\multirow{22}{*}{\textbf{Top 3}} & \multicolumn{1}{c}{Dataset} & \multicolumn{1}{c}{CF Model} & Baseline & + D   & + C   & + DC  & Baseline & + D                        & + C   & + DC   \\ 
\cmidrule{2-11}
                                 & \multirow{7}{*}{Canvas}     & EASE                               & 0.159    & 0.259 & 0.256 & \cg{0.262} & 0.16     & 0.259                      & 0.256 & \cg{0.262}  \\
                                 &                             & WRMF                               & 0.149    & 0.264 & 0.252 & \cg{\underline{0.275}} & 0.15     & 0.264                      & 0.252 & \cg{\underline{0.275}}  \\
                                 &                             & IKNN                               & 0.159    & 0.254 & 0.254 & \cg{0.263} & 0.16     & 0.254                      & 0.254 & \cg{0.263}  \\
                                 &                             & NMF                                & 0.072    & 0.173 & 0.205 & \cg{0.206} & 0.074    & 0.173                      & 0.205 & \cg{0.206}  \\
                                 &                             & SLIM                               & 0.152    & \cg{0.261} & 0.255 & 0.252 & 0.152    & \cg{0.261} & 0.255 & 0.252  \\
                                 &                             & SVD                                & 0.139    & \cg{0.242} & 0.217 & 0.231 & 0.141    & \cg{0.242}                      & 0.217 & 0.231  \\
                                 &                             & UKNN                               & 0.167    & \cg{0.259} & 0.249 & 0.247 & 0.166    & \cg{0.26}                       & 0.249 & 0.247  \\ 
\cmidrule{2-11}
                                 & \multirow{7}{*}{KDD}        & EASE                               & 0.429    & 0.628 & 0.596 & \cg{0.632} & 0.425    & 0.628                      & 0.596 & \cg{0.633}  \\
                                 &                             & WRMF                               & 0.296    & 0.535 & \cg{0.631} & 0.607 & 0.293    & 0.535                      & \cg{0.631} & 0.607  \\
                                 &                             & IKNN                               & 0.396    & 0.623 & 0.609 & \cg{0.645} & 0.392    & 0.623                      & 0.609 & \cg{0.645}  \\
                                 &                             & NMF                                & 0.32     & 0.589 & 0.556 & \cg{0.609} & 0.318    & 0.589                      & 0.555 & \cg{0.609}  \\
                                 &                             & SLIM                               & 0.438    & 0.62  & 0.614 & \cg{\underline{0.646}} & 0.435    & 0.621                      & 0.613 & \cg{\underline{0.646}}  \\
                                 &                             & SVD                                & 0.41     & 0.597 & 0.598 & \cg{0.633} & 0.407    & 0.597                      & 0.598 & \cg{0.632}  \\
                                 &                             & UKNN                               & 0.413    & 0.617 & 0.588 & \cg{0.633} & 0.417    & 0.589                      & \cg{0.633} & 0.632  \\ 
\cmidrule{2-11}
                                 & \multirow{7}{*}{XuetangX}   & EASE                               & 0.209    & 0.373 & 0.331 & \cg{0.392} & 0.244    & 0.373                      & 0.331 & \cg{0.392}  \\
                                 &                             & WRMF                               & 0.215    & 0.400 & 0.371 & \cg{0.411} & 0.253    & 0.4                        & 0.372 & \cg{0.411}  \\
                                 &                             & IKNN                               & 0.237    & 0.399 & 0.388 & \cg{0.413} & 0.234    & 0.399                      & 0.387 & \cg{0.413}  \\
                                 &                             & NMF                                & 0.158    & 0.359 & 0.362 & \cg{0.384} & 0.157    & 0.359                      & 0.362 & \cg{0.384} \\
                                 &                             & SLIM                               & 0.253    & 0.434 & 0.411 & \cg{\underline{0.441}} & 0.249    & 0.434                      & 0.411 & \cg{\underline{0.441}}  \\
                                 &                             & SVD                                & 0.22     & 0.373 & 0.377 & \cg{0.387} & 0.219    & 0.373                      & 0.377 & \cg{0.387}  \\
                                 &                             & UKNN                               & 0.243    & 0.407 & 0.379 & \cg{0.408} & 0.24     & 0.407                      & 0.379 & \cg{0.408} \\ 
\midrule
\multirow{21}{*}{\textbf{Top 5}} & \multirow{7}{*}{Canvas}     & EASE                               & 0.189    & \cg{0.307} & 0.304 & \cg{0.307} & 0.229    & \cg{0.308}                      & 0.304 & 0.306  \\
                                 &                             & WRMF                               & 0.183    & 0.313 & 0.317 & \cg{0.323} & 0.183    & 0.313                      & 0.317 & \cg{0.324}  \\
                                 &                             & IKNN                               & 0.19     & 0.307 & 0.303 & \cg{0.308} & 0.19     & \cg{0.307}                      & 0.303 & \cg{0.307}  \\
                                 &                             & NMF                                & 0.094    & 0.213 & \cg{0.239} & 0.234 & 0.095    & 0.213                      & \cg{0.239} & 0.234  \\
                                 &                             & SLIM                               & 0.183    & 0.309 & \cg{0.312} & 0.31  & 0.183    & 0.308                      & \cg{0.313} & 0.31   \\
                                 &                             & SVD                                & 0.167    & \cg{0.292} & 0.273 & 0.285 & 0.167    & \cg{0.292}                      & 0.273 & 0.285  \\
                                 &                             & UKNN                               & 0.198    & \cg{\underline{0.32}}  & 0.309 & 0.311 & 0.198    & \cg{\underline{0.32}}   & 0.309 & 0.311  \\ 
\cmidrule{2-11}
                                 & \multirow{7}{*}{KDD}        & EASE                               & 0.508    & \cg{0.653} & 0.623 & \cg{0.653} & 0.503    & \cg{0.653}                      & 0.621 & 0.652  \\
                                 &                             & WRMF                               & 0.373    & 0.573 & \cg{0.645} & 0.629 & 0.368    & 0.573                      & \cg{0.643} & 0.629  \\
                                 &                             & IKNN                               & 0.472    & 0.648 & 0.638 & \cg{\underline{0.666}} & 0.468    & 0.647                      & 0.636 & \cg{0.664}  \\
                                 &                             & NMF                                & 0.392    & 0.622 & 0.598 & \cg{0.636} & 0.388    & 0.622                      & 0.596 & \cg{0.635}  \\
                                 &                             & SLIM                               & 0.518    & 0.647 & 0.643 & \cg{\underline{0.666}} & 0.515    & 0.647                      & 0.641 & \cg{\underline{0.666}}  \\
                                 &                             & SVD                                & 0.487    & 0.628 & 0.623 & \cg{0.647} & 0.482    & 0.627                      & 0.622 & \cg{0.647}  \\
                                 &                             & UKNN                               & 0.489    & 0.646 & 0.621 & \cg{0.655} & 0.486    & 0.645                      & 0.62  & \cg{0.654}  \\ 
\cmidrule{2-11}
                                 & \multirow{7}{*}{XuetangX}   & EASE                               & 0.244    & 0.42  & 0.394 & \cg{0.433} & 0.241    & 0.418                      & 0.394 & \cg{0.432}  \\
                                 &                             & WRMF                               & 0.251    & 0.447 & 0.415 & \cg{0.448} & 0.248    & 0.446                      & 0.416 & \cg{0.448}  \\
                                 &                             & IKNN                               & 0.275    & 0.453 & 0.44  & \cg{0.458} & 0.272    & 0.451                      & 0.44  & \cg{0.458}  \\
                                 &                             & NMF                                & 0.195    & 0.399 & 0.404 & \cg{0.419} & 0.193    & 0.398                      & 0.404 & \cg{0.419}  \\
                                 &                             & SLIM                               & 0.297    & 0.483 & 0.464 & \cg{\underline{0.487}} & 0.293    & 0.482                      & 0.464 & \cg{\underline{0.487}}  \\
                                 &                             & SVD                                & 0.257    & 0.425 & \cg{0.431} & \cg{0.431} & 0.255    & 0.424                      & \cg{0.431} & \cg{0.431}  \\
                                 &                             & UKNN                               & 0.284    & 0.456 & 0.438 & \cg{0.457} & 0.281    & 0.454                      & 0.438 & \cg{0.457}  \\
\bottomrule
\end{tabular}}
\end{table*}

Beyond its superior performance compared to other competing methods, our proposed approach offers the added benefit of providing more clear explanations for recommendations. This allows us to determine the extent to which recommendations are influenced by enrollment likelihood (based on the course ranking in $L_{CF}$), time-to-completion (based on the course ranking in $L_{SA_{C}}$), or time-to-dropout (based on the course ranking in $L_{SA_{d}}$). For example, the explanation like: ``\textit{This course is recommended because learners similar to you have enrolled in these MOOCs"}, can be extended with ``\textit{We believe you will complete this course swiftly since we expect you finish this course X hours/days faster than the average student}, " or ``\textit{You will be more engaged with this course since you have X\% lower chance of dropout from the course relative to the average student}". Consequently, learners can effectively control various elements affecting recommendations and can disable any factors they deem irrelevant based on the provided explanations.

\section{Conclusion}
\label{sec:con}
Time-to-event information such as time-to-completion and time-to-dropout in MOOCs provides valuable insights into learners' preferences and needs. In this paper, we proposed modeling time-to-event in MOOCs—specifically, time-to-completion and time-to-dropout—using survival analysis methods. We sought to leverage these predictions to improve collaborative filtering recommender systems. This enhancement enables the recommender system to recommend courses that learners are both more likely to enroll in and complete quickly or stay engaged with for longer periods. As detailed in Section~\ref{sec:result}, our approach outperforms competing collaborative filtering-based recommender systems on three publicly available datasets, with better performance according to two variants of the NDCG measure.

There are two main directions for future work: (i) In this paper, we used survival analysis predictions to post-process the collaborative filtering-based recommendations. A promising direction for future work is to develop an ensemble model that can model user preferences and needs from different perspectives~\cite{gharahighehi2021ensemble}, or to model the problem as a multi-task learning problem to simultaneously learn two tasks: how likely a learner will enroll in a MOOC and how long it will take to complete the course or drop out from it. (ii) In this paper, we created two separate survival models for time-to-completion and time-to-dropout. A key assumption of most survival models is that all individuals will eventually experience the event of interest. In this context, a student who completes a course will never drop out of it when building a time-to-dropout model. We would like to investigate the merits of survival analysis methods with cured fraction information \cite{Amico2018} which remedies this by recognizing that some individuals will never experience the event of interest and explicitly models the probability of the event occurring inside the survival model.

\backmatter














\bibliography{sn-article}
\begin{appendices}

\section{Hyperparameter tuning}
\label{hyper}
Table~\ref{tab:hyper} provides the final tuned hyperparameters for the collaborative filtering and survival analysis models in our experiments. 

\begin{table*}[hbt!]
\caption{Selected hyperparameters}
\label{tab:hyper}
\scalebox{0.95}{
\begin{tabular}{lllllll}
\toprule
\multicolumn{4}{l}{}                                                    & \multicolumn{3}{c}{datasets} \\ \cmidrule{3-7} 
\multicolumn{2}{l}{}                                & parameter & range & XuetangX  & KDDCUP  & Canvas \\ \midrule
\multirow{15}{*}{CF} & UKNN    &\# neighbors & (20, 800)& 301 & 488 & 128 \\
                                          &         &shrink term& (0,1000)& 178& 907 & 8 \\
 \cmidrule{3-7}
                                          & IKNN    &\# neighbors&(20, 800)&  70   &  37       & 789        \\
                                          &         &shrink term&(0,1000)&   350 &  194       &   793     \\\cmidrule{3-7}
                                          & SVD     &\# latent features&(3, 50)&    5      &   3      &  8      \\\cmidrule{3-7}
                                          & NMF     &\# latent features & (10, 300) &    202       &   43      &     242   \\
                                          &         & L1\_ratio&(0.1,0.9)       &     0.214      &     0.846    &     0.232   \\\cmidrule{3-7}
                                          & WRMF    & epochs & (10,200)      &   217        &    200  &   10     \\
                                          &         & \# latent features &  (10,100)     &   21        &    45     &  43      \\
                                          &         &  regularization         & (1e-5, 1e-1)&   0.008        &  0.097       &0.093         \\\cmidrule{3-7}
                                          & EASE    & l2\_norm & (1e0, 1e7)      & 95109 & 2540 &  9325573      \\\cmidrule{3-7}
                                          & SLIM    & topk & (50, 600)      &   380    &    486    & 321  \\
                                          &         & l1\_norm  & (1e-5,1.0)  &  0.0002     &  0.593     & 0.039   \\
                                          &         & l2\_norm & (1e-3, 1.0) &  0.310  &   0.006      &    0.179 \\ \midrule
\multirow{10}{*}{$SA_{D}$}                       & CoxNet  & alpha          &   (0,1)    &  0.0039         &   0.00941      &    0.335    \\ \cmidrule{3-7}
                                          & RSF     &   n\_estimators         &  (25,100)     & 100          & 100        & 80       \\
                                          &         &  min\_samples\_leaf         &  (10,20)     & 13          & 12        & 18       \\
                                          &         &  min\_samples\_split         &  (10,20)     & 11          & 18        & 20       \\
                                          &         &  max\_depth         &  (2,12)     & 12          & 12       & 9       \\\cmidrule{3-7}
                                          & XGBoost &   Learning Rate       &  (0.1,1)     &     0.2777      &  0.4124       &    0.1247    \\
                                          &         &  n\_estimators         &  (25,200)     & 184          & 121        & 199       \\
                                          &         &  min\_samples\_leaf         &  (5,20)     & 16          & 16        & 17       \\
                                          &         &  min\_samples\_split         &  (5,20)     & 6          & 6        & 16       \\
                                          &         &  max\_depth         &  (2,20)     & 7          & 16       & 12       \\ \midrule
\multirow{10}{*}{$SA_{C}$}                       & CoxNet  & alpha          &   (0,1)    &     0.0050      &   0.0639      &     0.236   \\ \cmidrule{3-7}
                                          & RSF     &   n\_estimators         &  (25,100)     & 78          & 75        & 98       \\
                                          &         &  min\_samples\_leaf         &  (10,20)     & 17          & 13        & 19       \\
                                          &         &  min\_samples\_split         &  (10,20)     & 13          & 17        & 15       \\
                                          &         &  max\_depth         &  (2,12)     & 12          & 6       & 7       \\\cmidrule{3-7}
                                          & XGBoost &   Learning Rate       &  (0.1,1)     &     0.3835      &  0.1495       &    0.1117    \\
                                          &         &  n\_estimators         &  (25,200)     & 173          & 124        & 153       \\
                                          &         &  min\_samples\_leaf         &  (5,20)     & 10          & 8        & 10       \\
                                          &         &  min\_samples\_split         &  (5,20)     & 12          & 6        & 12       \\
                                          &         &  max\_depth         &  (2,20)     & 10          & 12       & 10       \\
                                          \bottomrule
\end{tabular}}
\end{table*}

\end{appendices}



\end{document}